\documentstyle[preprint,aps]{revtex}
\begin{document}
\preprint{\tt SNUTP 95-027}
\draft
\title{Renormalization group analysis of \\
 the anisotropic Kardar-Parisi-Zhang equation \\
 with spatially correlated noise \\}
\author{H. Jeong$^{1}$, B. Kahng$^{2}$, and D. Kim$^{1}$\\}
\address{$^{1}$ Center for Theoretical Physics and Department of Physics, \\
Seoul National University, Seoul 151-742, Korea \\
$^{2}$ Department of Physics, Kon-Kuk University, Seoul 133-701, Korea \\}
\maketitle

\begin{abstract}
We analyze the anisotropic Kardar-Parisi-Zhang equation in general
substrate dimensions $d'$ with spatially correlated noise,
$\langle\tilde \eta({\bf{k}},\omega)\rangle=0$ and
$\langle\tilde \eta({\bf{k}},\omega) \tilde \eta({\bf{k}}',\omega')\rangle$
$=2D(k)\delta^{d'}({\bf{k}}+{\bf{k}}')\delta(\omega+\omega')$
where $D(k)=D_0+D_{\rho}k^{-2\rho}$, using the dynamic renormalization
group (RG) method. When the signs of the nonlinear terms
in parallel and perpendicular directions are opposite, a novel finite stable
fixed point is found for $d'< d'_c \equiv 2+2\rho$
within one-loop order. The roughening exponent $\alpha$ and the dynamic
exponent $z$  associated with the stable fixed point are determined as
$\alpha={2 \over 3}\big(\rho-{{d'-2}\over 2}\big)$, and $z=2-\alpha$.
For $d' > d'_c$, the RG transformations flow to the fixed
point of the weak-coupling limit, so that the dynamic exponent becomes $z=2$.
\end{abstract}

Recently, there have been many studies in the field of
nonequilibrium surface growth.
A number of discrete models and continuous equations
for the surface growth phenomena have been introduced
and studied~\cite{review}.
One of the most interesting features in the nonequilibrium surface growth
is the nontrivial dynamic scaling behavior of the interface
width, i.e.,
\begin{equation}
W(L,t) = \langle {1 \over L^{d'}}\sum_i(h_i - {\bar h})^2 \rangle^{1/2}
\sim L^{\alpha}f(t/L^z),
\end{equation}
where $h_i$ is the height of site $i$ on the substrate. $\bar h$, $L$, and
$d'$ denote the mean height, system size, and the substrate dimension,
respectively. The symbol $\langle \cdots \rangle$ stands
for the statistical average.
The scaling function $f(x)$ approaches to a constant for $x \gg 1$, and
$f(x) \sim x^{\beta}$ for $x \ll 1$ with $z=\alpha/\beta$. The
exponents $\alpha$, $\beta$ and $z$ are called the roughness,
the growth, and the dynamic exponent, respectively. \\

The simplest nonlinear dynamic equation describing a growing surface
with lateral growth effect was introduced and studied by Kardar, Parisi,
and Zhang (KPZ)~\cite{kpz}, which is given by
\vspace{.1in}
\begin{equation}
{{\partial h({\bf{x}},t)}\over {\partial t}}=\nu \nabla^2 h
+{\lambda \over 2} (\nabla h)^2+\eta({\bf{x}},t).
\end{equation}
\vspace{.1in}
The noise $\eta({\bf{x}},t)$ is, in general, correlated
spatially and temporally,
$\langle \tilde \eta({\bf{k}}, \omega) \rangle=0$, and
\begin{equation}
\langle \tilde \eta({\bf{k}}, \omega) \tilde \eta({\bf{k}}', \omega') \rangle =
2D({\bf{k}},\omega)\delta^{d'}({\bf{k}}+{\bf{k}}') \delta(\omega+\omega'), \\
\end{equation}
where $\tilde \eta({\bf{k}}, \omega)$ is the Fourier transform of
the noise $\eta({\bf{x}},t)$.
When the noise is correlated only spatially, an interesting form of
$D({\bf{k}},\omega)$ can be written as $D(k)=D_0+D_{\rho}k^{-2
\rho}$~\cite{medina}.
Here, the $D_0$ term is needed to make the RG transformation closed.
The white noise corresponds to the limiting case of $\rho=0$.
When $\rho > 0$, the noise is correlated in space with a power law. It is
useful to remind that the KPZ equation is invariant under the Galilean
transformation,
which yields the scaling relation, $\alpha+z=2$~\cite{kpz}. \\

Recently the anisotropic KPZ (AKPZ) equation was introduced to
study the surface growth on the vicinal substrate~\cite{wolf}.
Besides, the AKPZ equation can also be applied to various
physical growth problems such as
the ion-sputtered surface growth~\cite{cb}, and
the surface growth on the reconstructed surface structure~\cite{vven},
$\sl etc$. The AKPZ equation is written as
\begin{equation}
{{\partial h({\bf{x}},t)}\over {\partial t}}=\nu_{\perp} \nabla_{\perp}^2 h +
\nu_{\parallel} \nabla_{\parallel}^2 h + {\lambda_{\perp} \over 2}
(\nabla_{\perp} h)^2 +
{\lambda_{\parallel} \over 2}(\nabla_{\parallel} h)^2+\eta({\bf{x}},t),
\end{equation}
where $\nabla_{\perp} (\nabla_{\parallel})$ is the gradient along the
perpendicular (parallel) directions.
The anisotropy means $r_{\nu} \equiv \nu_{\parallel} / \nu_{\perp} \ne 1$ and
$r_\lambda \equiv \lambda_{\parallel} / \lambda_{\perp} \ne 1$.
Here, we consider only the case of positive $\nu_{\perp}$ and $\nu_{\parallel}$
for stable surfaces.
The AKPZ equation was studied by Wolf for the case of white noise
in 2+1 dimensions using the dynamic renormalization group (RG)
method~\cite{wolf}.
He found that when the signs of $\lambda$'s are opposite,
the nonlinear terms turn out to be irrelevant under the RG transformation.
As a result, the AKPZ equation with opposite signs of $\lambda$'s belongs
to the weak coupling limit, the Edwards-Wilkinson (EW) universality
 class~\cite{ew}.
A stochastic model associated with this weak coupling behavior of the AKPZ
equation in 2+1 dimensions was introduced by the present authors~\cite {jkk}
as a generalization of the Toom model~\cite{toom}.
By the simulations, we confirmed that the height fluctuations width
increases logarithmically both with time before saturation,
 and with system size after saturation,
which is the signature of the EW universality class. \\

In this paper, we extend the study of the AKPZ equation to the case
of spatially correlated noise,
$D(k)=D_0+D_{\rho}k^{-2 \rho}$, in general substrate dimensions $d'$,
using the dynamic RG method. Thus the previous study by Wolf
corresponds to the limiting case, $\rho=0$ and $d'=2$, of the current study.
By this extension, we find that when the signs of the nonlinear coupling
coefficients are opposite, a new ``stable'' fixed point of the
RG flow exists within the one-loop order in the ``off-axis region'' of
the parameter space $(U_0,U_{\rho})$ provided $d'< d'_c \equiv 2+2\rho$.
Here $U_0$ and $U_{\rho}$ are dimensionless effective coupling constants, $U_0
\equiv K_{d'-1}{{ D_0 \lambda_{\perp}^2} / \nu_{\perp}^3}$ and
$U_{\rho} \equiv K_{d'-1}{{ D_{\rho} \lambda_{\perp}^2} / \nu_{\perp}^3}$,
where $K_{d'}=S_{d'}/(2\pi)^{d'}$, with $S_{d'}$ the surface area of a unit
sphere in $d'$ dimensions.
Moreover, the dynamic exponent $z$ associated with the fixed point is
determined exactly as $z=2-{2 \over 3}(\rho - {{d'-2}\over 2})$.
Thus, even at $d'=2$, the exponents $\alpha$ and $z$ are exactly obtained as
$z=2-{2\rho \over 3}$ and $\alpha={2\rho \over 3}$ for finite $\rho$.
This result would be interesting,  because it is rare to know
the dynamic exponent exactly at $d'=2$ in other surface growth problems.
As the dimension $d'$ approaches $d_c'$ for finite $\rho$,
the stable fixed point moves to the origin, $(U_0^*, U_{\rho}^*) = (0,0)$,
and the nonlinear terms become irrelevant for $d'> d'_c$.
Conversely, as the power $\rho$ approaches zero at $d'=2$,
the fixed point shifts towards the origin, and the weak coupling behavior
derived by Wolf is recovered.
On the other hand, when the signs of the nonlinear terms are
the same, the behavior of the RG flows is similar to the one for the
isotropic case with the correlated noise studied by Medina $\sl et$
$\sl al$.~\cite{medina}, so that we will not consider this case further. \\

The equation we study here is the AKPZ equation, Eq.~(4), in general
$d'$ dimensions, among which one dimension is assigned to
the parallel dimension on the substrate, and the remaining $d'-1$
dimensions to the perpendicular dimensions.
The calculations of the RG transformation were performed by
combining the two methods, one of which was introduced
 by Medina $\sl et$ $\sl al.$ to study the isotropic case with the
correlated noise, and the other by Wolf to study the anisotropic case with the
white noise. The steps of the RG transformation are following.
First, the anisotropic exponent $\chi$ is introduced to relate
the two characteristic length scales as
$\xi_{\parallel} \sim \xi_{\perp}^{\chi}$, where $\xi_{\parallel}$
and $\xi_{\perp}$ are the characteristic length scales in parallel and
perpendicular directions on the substrate, respectively.
Next, the coarse-graining transformation is performed within the one-loop order
by integrating out the fluctuations of heights within small
length scales, which correspond to the wavevectors,
$e^{-l}\pi/a \le |k_{\perp}| \le \pi/a$
and $e^{-l\chi}\pi/a \le |k_{\parallel}| \le \pi/a$, ($a$ is
the lattice constant) for the parameters,
$\nu_{\perp}$, $\nu_{\parallel}$, $\lambda_{\perp}$, $\lambda_{\parallel}$,
and the effective coupling constants, $U_0$, and
$U_{\rho}$. Through this coarse-graining process, the
$\lambda$'s remain unchanged, which is due to the invariance of
the AKPZ equation under the Galilean transformation,
\begin{equation}
x_i \rightarrow x_i + \lambda_i \epsilon t, \phantom{xxxxxxxx}
h \rightarrow h+\epsilon \sum_i x_i-\sum_i\lambda_i \epsilon^2 t /2,
\end{equation}
where $i$ denotes $\perp$ or $\parallel$, and $\epsilon$ is
an infinitesimal angle of tilt. As a final step,
the rescalings are performed as $x_{\perp} \rightarrow e^{l} x_{\perp}$,
$x_{\parallel} \rightarrow e^{l\chi} x_{\parallel}$,
$h \rightarrow e^{l \alpha} h$, $t \rightarrow e^{lz} t$.
Under the rescalings, one has $\lambda_{\perp} \rightarrow
e^{l(\alpha+z-2)}\lambda_{\perp}$ and $r_{\lambda} \rightarrow
e^{2l(1-\chi)}r_{\lambda}$. From the scale invariance,
the scaling relation, $\alpha+z=2$, and $\chi=1$ are obtained,
provided that $\lambda_{\perp} \ne 0$ and $r_{\lambda} \ne 0$.
The case of $\lambda_{\perp} \ne 0$ and $r_{\lambda} \ne 0$
is much more realistic than the case that one of the $\lambda$'s is
zero. Fortunately for $\chi=1$, the recursion relations are explicitly
calculable,
 but they could not be done for $\chi \ne 1$. We find that the RG recursion
relations for $\chi = 1$
 are as follows.
\begin{eqnarray}
{{\partial \nu_{\perp}} \over {\partial l}}&=&\nu_{\perp} \Big[z-2
-(U_0+U_{\rho})(A-{{2 C} \over {d'-1}})+U_\rho{{2\rho B} \over {d'-1}}
\Big],  \\
{{\partial r_{\nu} } \over {\partial l}} &=& r_{\nu} (U_0+U_{\rho})
\Big[(1-{r_\lambda \over r_{\nu}}) A -{ {2 C} \over {d'-1}} + 2 r_\lambda
E\Big]+
2 \rho r_{\nu} U_{\rho}\Big[\Big(- { B \over {d'-1}}+{r_\lambda F \over
{r_{\nu}}}
\Big ) \Big], \\
{{\partial U_0} \over {\partial l}}&=&U_0(2-d')+U_0^2
(G+3 A-{{6C}\over {d'-1}}) +U_0 U_{\rho}\big(3A-{{6C+6\rho B}\over{d'-1}}+2G
\big)+U_{\rho}^2 G, \\
{{\partial U_{\rho}} \over
{\partial l}}&=& U_{\rho} \Big[(2-d'+2\rho)+3 (U_0+U_{\rho})
(A -{{2 C}\over {d'-1}})-U_{\rho} {{6 \rho B} \over {d'-1}}\Big].
\end{eqnarray}
Here $A,B,C,E,F$, and $G$ are defined as
$A\equiv I(0,1,{{d'-2}\over 2},2)$, $B\equiv I(0,1,{d'\over 2}, 2)$,
$C\equiv I(0,1,{{d'-2}\over 2},3)$, $E\equiv I(2,1,{{d'-2}\over 2},3)$,
$F\equiv I(2,1,{d'\over 2},2)$, and $G\equiv I(0,2,{{d'-2}\over 2},3)$,
respectively, where \\
\begin{equation}
I(\alpha, \beta, \gamma, \delta) \equiv \int_0^{\infty} dy
{{y^{\alpha}(1+r_{\lambda} y^2)^{\beta}}\over
{(1+y^2)^{\gamma}(1+r_{\nu} y^2)^{\delta}}}.
\end{equation}
Thus, $A,B,C,E,F,G$ are functions of the variables, $d', r_{\nu}$, and
$r_\lambda$. \\

Before we analyze the RG recursion relations in detail, it is interesting to
note that Eqs.~(6) and (9) have the similar expressions. To make parameters
unchanged, if we set ${\partial r_{\nu}} / {\partial l}=0$ and ${\partial
U_{\rho}} / {\partial l} =0$,
we can easily get the relation, $z-2=-(2-d'+2\rho)/3$. So the dynamical
exponent can be written as
\begin{equation}
z=2-{2\over3}(\rho-{{d'-2}\over2}), \phantom{xxxxxxxxx}
\alpha={2\over3}(\rho-{{d'-2}\over2}),
\end{equation}
provided $\nu_{\perp} \neq 0$ and $U_{\rho} \neq 0$. Thus, if there exists a
stable fixed point at finite values of $\nu_{\perp}$ and $U_{\rho}$, we then
immediately have the expressions for $z$ and $\alpha$. \\

In order to find the stable fixed point of
the recursion relations,
it is of help to recall the RG flow for the case
of the white noise, $\rho=0$ at $d'=2$~\cite{wolf}. For this case,
there exist the fixed points $r_{\nu}^*=\pm r_\lambda$, which were obtained
by setting ${{\partial r_{\nu}} / {\partial l}}=0$.
For $r_{\nu}^* =r_\lambda$, the fixed point of the effective coupling
constant, $U \equiv D\lambda^2 /\nu^3$ at zero,
is unstable, while it is stable for $r_{\nu}^* =-r_\lambda$.
Thus for $r_\lambda < 0$, the nonlinear terms of the AKPZ equation become
irrelevant even though they are finite, and the AKPZ equation belongs to the EW
universality class.
Following the similar steps, we have analyzed the recursion relations
for the case of correlated noise with $d'=2$ first. When $d'=2$,
Eq.~(7) becomes explicitly as
\begin{equation}
{{\partial r_{\nu}} \over {\partial l}}={{\pi (r_{\nu} + r_\lambda)}\over {8
(1+\sqrt{r_{\nu}})^2 r_{\nu}^{3 \over 2}}} \Big[
(U_0 + U_{\rho})(1+\sqrt{r_{\nu}})^2 (r_\lambda - r_{\nu}) +
4 \rho U_{\rho} (r_\lambda + 2 r_\lambda \sqrt{r_{\nu}} - 2 r_{\nu}^{3 \over 2}
- r_{\nu}^2)
\Big].
\end{equation}
Thus, $r_{\nu}^*=-r_\lambda$ is the solution of
${\partial r_{\nu}} / {\partial l}=0$ even for the case of correlated noise.
Solving ${\partial U_0} / {\partial l} =0$ and
${\partial U_{\rho}} / {\partial l} =0$ with $r_{\nu}^*=-r_\lambda$ in
Eqs.~(8) and (9), we then find two fixed points other than that of
$U_0^*=U_{\rho}^*=0$ which is unstable for $\rho > 0$ anyway. The stability
analysis shows that,
of the two new fixed points, one is unstable and the other is stable for
all $\rho>0$ and $r_\lambda<0$. The stable
fixed point is located at
\begin{equation}
   U_\rho^* = \frac{4\rho \sqrt{r_\nu^*}}{\pi}
	\left(	\frac{-\zeta-9+3\sqrt{\zeta^2+2\zeta+9}}{\zeta^2}
\right),		\end{equation}
\begin{equation}
U_0^*=\frac{4\rho \sqrt{r_\nu^*}}{\pi} \left(
 \frac{\zeta^2+4\zeta+9-(\zeta+3)\sqrt{\zeta^2+2\zeta+9}}{\zeta^2} \right) ,
		\end{equation}
in the one-loop approximation where
\begin{equation}
\zeta=\frac{12\rho\sqrt{r_\nu^*}}{(1+\sqrt{r_\nu^*})^2}.
\end{equation}
Thus, we indeed have a stable fixed point in the `off-axis' region and
hence the dynamic exponents are given by Eq.~(11) with $d'=2$.
As $\rho$ decreases to zero, the stable fixed
point moves to ($U_0^*$, $U_{\rho}^*$)=(0,0), and the case of
white noise is recovered.
As mentioned before, the recursion relations for $\chi \ne 1$ could not
be derived explicitly. Thus we consider the case of $\chi \ne 1$ by
extrapolating the analysis for $\chi = 1$ with the recursion relations,
Eqs.~(6-9) in the following way.
Since $\chi \ne 1$ means $r_\lambda =0$, we examine the behavior of
the fixed point by taking the limit $r_\lambda \rightarrow 0^-$ for fixed
$\rho$.
As can be seen in Eq.~(13) and (14), the stable fixed point moves to the origin
in the parameter
space as $r_\lambda$ approaches to $0^-$, which suggests that the case of
$\chi \ne 1$ belongs to the weak coupling limit.

For general dimensions $d'$, the explicit expressions of the
recursion relations are more complicated.
Thus, it is difficult to derive the analytic formula for $r_{\nu}^*$ from
${\partial r_{\nu}} / {\partial l} =0$. Instead, the stable
fixed point, ($r_{\nu}^*$, $U_0^*$, $U_{\rho}^*$), is found numerically by
iterating
the RG transformation.
We find that the stable fixed point exists in the off-axis region
of the parameter space for $d' < d'_c \equiv 2+2\rho$, and it locates
at $U_0^*=U_{\rho}^*=0$ for $d' > d'_c$.
This is in accord with Eq.~(9) which shows that the fixed point at
$U_0^*=U_\rho^*=0$ is unstable for $2-d'+2 \rho>0$. Fig.\ \ref{fig1} shows a
typical RG flow diagram projected onto the ($U_\rho$,$U_0$) plane
at $d'=2.5$. The existence of the off-axis fixed point implies that
the nontrivial exponents $z$ and $\alpha$ are given by Eq.~(11) for $d <
d'_c$.\\

The analytic formula, Eq.~(11) is derived from the one-loop
approximation RG transformation. However, it is in fact exact to all orders.
The invariance
of the noise spectrum $D(k)$ under the scale change,
$x \rightarrow bx$, $t \rightarrow b^z t$,
and $h \rightarrow b^{\alpha}h$ gives the exact exponent identity
$2\rho-d'-2\alpha+z=0$ as argued in \cite{medina}. This together with
the relation $\alpha+z=2$ gives Eq.~(11). The difference between the case
treated in \cite{medina} and in this work is the range of applicability
of Eq.~(11). For the isotropic case, the formula is valid for $d'<3/2$ and
$\rho_{\min} < \rho < \rho_{\max}$ with approximately known $\rho_{\max}$ and
$\rho_{\min}$.
However, for the anisotropic case, the formula for the dynamic exponent
is valid for $d' < d'_c= 2+2 \rho $ and $\rho >0$.
Hence the dynamic exponent
$z$ in $d'=2$ is nontrivial for finite $\rho$.
Since the exponents $\alpha$ and $z$ are nontrivial in $d'=2$,
it would be interesting to introduce a stochastic model associated
with the AKPZ equation having the alternative signs of $\lambda$'s
for correlated noise. A possible candidate for such a stochastic model
would be a generalization of the 2+1 dimensional Toom model introduced
for the white noise~\cite{jkk} by replacing the random deposition
by the spatially correlated deposition following the L\'evy flight
distribution. This conjecture is based on the analogy to the isotropic case,
where the correlated noise was simulated by the correlated
particle deposition following the L\'evy flight distribution~\cite{mj}.
The simulations of the model is underway~\cite{future}. \\

In summary, we have considered the anisotropic KPZ equation with
the spatially correlated noise, using the one-loop dynamic RG transformation.
In this study, the case that the signs of the nonlinear
terms are opposite is interesting for both of white and correlated
noises. Unlike the case of white noise, the stable fixed
point exists in the ``off-axis" region of the parameter space of
the effective coupling constants $U_0$ and $U_{\rho}$
if $d' < d'_c=2+2\rho$ for the case of correlated noise. From the stable fixed
point, the dynamic exponent $z$ is
determined as $z=2-{2 \over 3}
(\rho-{{d'-2}\over  2})$ and the roughening exponent
$\alpha=2-z$. Therefore, the nontrivial values of the exponents $\alpha, z$ are
present
in $d'=2$ for finite $\rho$. For $d' > d'_c$, the stable fixed point is located
at the origin in the parameter space,
and the AKPZ equation belongs to the Edwards-Wilkinson's
universality class. \\

This work is supported in part by the KOSEF under the contract
No 941-0200-006-2 and through the SRC program of SNU-CTP,
and in part by the BSRI program of Ministry of Education, Korea. \\

%\newpage

\begin{figure}
\caption{ Plot of the flow diagram of the RG transformation
in the parameter space of $U_{\rho}$ and $U_0$ for
$d'=2.5$, $\rho=0.5$ and $r_\lambda=-2.5$.}
\label{fig1}
\end{figure}
\end{document}